\documentclass[aps,pre,preprint]{revtex4-1}

\usepackage{graphicx}
\usepackage{amsmath,amssymb}
\usepackage[noload=abbr]{siunitx}

\newcommand{\ks}{\ensuremath{k_{s}}}
\newcommand{\ki}{\ensuremath{k_{i}}}
\newcommand{\kid}{\ensuremath{\kappa_{i}}}

\usepackage{hyperref}

\begin{document}

\title{Role of dynamic capsomere supply for viral capsid self-assembly}

\author{Marvin A. Boettcher}
\thanks{These authors contributed equally.}
\author{Heinrich C. R. Klein}
\thanks{These authors contributed equally.}
\author{Ulrich S. Schwarz}
\affiliation{Institute for Theoretical Physics and BioQuant, Heidelberg University, Heidelberg, Germany}

\begin{abstract}
Many viruses rely on the self-assembly of their capsids to protect and transport their genomic material. 
For many viral systems, in particular for human viruses like hepatitis B, adeno or human immunodeficiency
virus, that lead to persistent infections, capsomeres are continuously produced in the cytoplasm of the host cell while
completed capsids exit the cell for a new round of infection.
Here we use coarse-grained Brownian dynamics simulations of a generic 
patchy particle model to elucidate the role of the dynamic supply of capsomeres for the reversible self-assembly of empty T1 icosahedral virus capsids. We find that for high rates of capsomere influx only a narrow range of bond strengths exists for which a steady state of continuous capsid production is possible.
For bond strengths smaller and larger than this optimal value, the reaction volume becomes crowded by small and large intermediates, respectively. 
For lower rates of capsomere influx a broader range of bond strengths exists for which a steady state of
continuous capsid production is established, although now the production rate of capsids is smaller. 
Thus our simulations suggest that the importance of an optimal bond strength for viral capsid assembly typical for in vitro conditions
can be reduced by the dynamic influx of capsomeres in a cellular environment.
\end{abstract}

\maketitle

\section{Introduction}
Viruses are experts on the inner working of cells and their investigation has contributed strongly to our understanding
of the physical principles at work in biological systems. In particular, the study of viruses
has demonstrated the amazing power of biological self-assembly in an experimentally and theoretically accessible system. 
In their physiological context, viruses rely on the molecular machinery of their host
to reproduce both their genomic material and the protein capsid usually encapsulating it~\cite{Roos2010,Cann2001}.
Typically the capsid  is assembled from many copies of only a few different capsid proteins
and shows icosahedral or helical symmetry \cite{Rossmann1989,Caspar1956,Franklin1955,Crick1956,Caspar1962}.
The elementary assembly blocks for the assembly of a capsid are termed \textit{capsomeres}. They can either consist of single capsid proteins or of preassembled sets of capsid proteins.
For many viruses the formation of the capsid can be reproduced in vitro \cite{Fraenkel-Conrat1955,Finch1968}.
The dynamics of in vitro assembly has been analyzed using light and small-angle X-ray scattering techniques~\cite{Zlotnick1999,Zlotnick2000,Casini2004,Kler2012}.
While capsid assembly of viruses with single-stranded genomic material often requires the presence of the 
genomic material as an 'electrostatic glue' \cite{Bruinsma2006,Roos2010}, this is typically not possible for
double-stranded genomic material due to its larger bending stiffness. These viruses typically assemble
their capsid without the genomic material, which is then inserted into the capsid by a motor \cite{Sun2010,Roos2010}.

Despite the plethora of known capsid structures~\cite{Carrillo-Tripp2009}, the dynamic assembly process of the virus shell is still far from being fully understood. As experimental possibilities for detailed
monitoring of the assembly dynamics are limited, modeling can significantly help to increase our understanding of the mechanisms that govern the assembly process.
In the past various techniques
ranging from coarse-grained molecular dynamics or Brownian dynamics simulations~\cite{Schwartz1998,Rapaport1999,Rapaport2004,Hagan2006,Freddolino2006,Nguyen2007,Rapaport2008,Rapaport2012,Baschek2012}
through Monte Carlo simulations~\cite{Wilber2007,Wilber2009,Johnston2010} and discrete stochastic approaches~\cite{Zhang2005a,Zhang2006,Hemberg2006,Keef2006,Sweeney2008}
to thermodynamic descriptions \cite{Zlotnick1994,Zlotnick1999,Endres2002,Katen2009} have been used to elucidate different aspects of the assembly process. 
For example, the influence of capsomere shape~\cite{Rapaport2004,Rapaport2012} or the emergence of polymorphic structures~\cite{Nguyen2008,Elrad2008,Nguyen2009} have been investigated
with such theoretical approaches. The different techniques used to study virus capsid assembly have been recently reviewed by Hagan~\cite{Hagan2014}.

Several studies have shown that the successful assembly of complete capsids starting from a fixed number of assembly subunits requires intermediate (or even optimal) bond strengths~\cite{Endres2002,Hagan2006,Nguyen2007,Rapaport2008,Katen2009,Johnston2010,Zlotnick2011,Hagan2011,Baschek2012}.
At low bond strengths (or high temperature) the formation of large clusters is suppressed. 
At high bond strengths (or low temperature), by contrast, the simultaneous formation of stable assembly intermediates, which cannot recombine into complete capsids, or the formation of  large, misassembled structures with non-native binding interactions can prevent the formation of complete capsids. In this case the system becomes kinetically trapped. 

While in vitro studies and computer simulations usually work with a finite initial number of capsomeres that are increasingly used up during the capsid assembly process, in the physiological context of the cell the capsomeres are produced in a continuous fashion.  
In a recent stochastic simulation study, it has been shown for the case of genome-stabilized capsids (as for example
for the bacteriophage MS2) that this 'protein ramp' can make virus assembly very robust against kinetic trapping \cite{Dykeman2014}.
While bacteriophages typically kill the host cell to start a new round of infection, many animal and plant viruses
tend to follow  strategies that keep the host cell alive at least for a certain period of time. 
This is especially true for human viruses such as hepatitis B virus (HBV), human adeno virus (HAdV) or human immunodeficiency virus (HIV),
which lead to persistent infections. Recently the role of dynamic protein supply for viral capsid assembly has been studied
from a systems level perspective using a kinetic
gene expression model with exponential protein production and a master equation
for capsid assembly~\cite{Zhdanov2014}. However, the effect
of continuous protein production has not been studied yet in a spatial model. For empty capsid assembly, it has been shown
earlier with Brownian dynamics that steady states exist in which the removal of large clusters 
is balanced by the reinsertion of the corresponding monomers~\cite{Hagan2011}.  However,
the effect of the capsomere supply rate on these steady states has not been studied yet.

Inspired by the notion of continuous virus production, we use coarse-grained Brownian dynamics simulations to investigate the assembly
of empty T1 capsids in the presence of a dynamic capsomere supply. In order to avoid crowding of the reaction volume by
large clusters and motivated by the exit of completed virions from the cell by budding or exocytosis~\cite{Freed2004,Morita2004,Saksena2006,Mukhopadhyay2005}, 
capsids are removed from the simulation in the moment that they are completed.
Our simulations suggest that there exists a certain range of bond strengths in which the influx of new capsomeres is balanced by capsid removal. This steady-state region is surrounded by parameter regions in which successful capsid assembly is prohibited by crowding of the reaction volume; depending on influx rate and bond strength, crowding is observed either with small or large intermediates.
Our main result is that the favorable region for continuous virus production becomes larger for lower influx rates.
Our work identifies essential limits of viral self-assembly in a dynamic context and suggests that bond strength has to be less fine-tuned in a cellular context than in in vitro experiments. 

\section{Methods}
\begin{figure*}
    \includegraphics{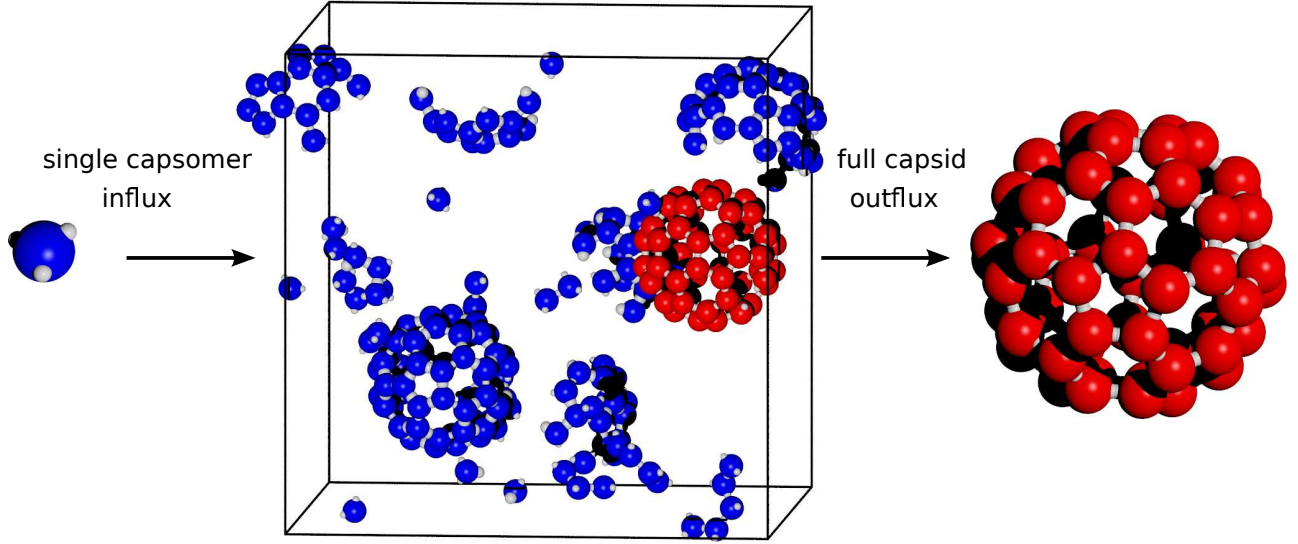}
    \caption{Schematic illustration of our simulations. Capsomeres are randomly inserted into the simulation volume with a rate $\ki$ while completed capsids are removed immediately (including the intermediates they might contain inside). This simple model mimics the situation in human cells with persistent infections
in which capsomeres are continuously produced by translation and completed capsids leave by budding or exocytosis.}
        \label{fig:Model_explanation}
\end{figure*}
To investigate the effect of a continuous influx of capsomeres on virus assembly, we use an efficient coarse-grained Brownian dynamics approach, which has previously been used to study the effect of reactivity switching during the assembly process~\cite{Baschek2012}. 
Here we consider a T1 capsid which is composed of 60 identical capsomeres \cite{Caspar1962}.
Each capsomere is described by a hard sphere which is equipped with spherical patches (see Fig.~\ref{fig:Model_explanation}). 
The spatial arrangement of the patches reflects the capsid geometry according to the local rules scheme developed by Berger et al.~\cite{Berger1994,Schwartz1998}. 
All assembly intermediates formed during the assembly process are treated as rigid objects and are propagated according to their translational and rotational diffusive properties~\cite{Schluttig2008,Schluttig2009}, which are evaluated on-the-fly upon their formation~\cite{Carrasco1999,Schluttig2009}.

If an overlap between two complementary patches is realized by diffusion, a bond is formed with the probability $P_\text{react}=k_a \Delta t \ll 1$. 
This bond can be established either between two unconnected clusters (inter-bond) or in an already connected cluster (intra-bond). 
Otherwise we only consider one type of bond to keep the number of parameters small.
We assume that bond formation is achieved by strong local forces such that this process is very fast on the time scale of our simulations.
Therefore upon formation of an inter-bond, the clusters instantaneously assume the correct relative position and orientation for the
assembly of the capsid unless the necessary reorientation results in a steric overlap either between the two merging clusters or with other clusters. 
We note that our approach based on patchy particles and local rules does not allow us to study the formation of
aberrant cluster with non-native interactions \cite{Nguyen2007,Hagan2006,Hagan2011}. Because we do not consider any forces, our approach also
does not allow us to investigate strained capsids, as possible in molecular dynamics or Brownian dynamics
simulations with potentials.

In order to study reversible dynamics, every existing bond can also dissociate with the probability $P_\text{dissoc}=k_d \Delta t \ll 1$.  
If bond dissociation results in two unconnected clusters, they are positioned relative to each other according to a computational scheme which ensures detailed balance in order to prevent additional, non-physical driving forces for the self-assembly~\cite{Klein2014}.

Intra-bond formation in an already connected cluster leads to an additional stabilization of the cluster as closed loops are formed, in which every capsomere is connected to at least two neighboring capsomeres. 
For such a loop structure to break apart it is necessary that all bonds have to be in the open state simultaneously. 
The formation/dissociation of an intra-bond does not affect the structure of the complex. 
The energy gain by bond formation is related to the microscopy rates by $E=-k_B T \ln (k_a / k_d)$, where $k_B$ is the Boltzmann constant~\cite{Klein2014}. 

To study the effect of a continuous supply of capsomeres on the assembly dynamics, we introduce the influx rate $\ki$.
We place a new capsomere in the simulation volume in each time step with the probability $p_\text{in}=\ki \Delta t \ll 1$. Position and orientation of the new capsomere are randomly chosen with the constraint that no steric overlap with existing clusters is created. Inspired by virus exit strategies like budding or exocytosis, which in principle can lead to a steady state of virus production, complete capsids are removed immediately upon formation together with all intermediates inside the capsid (compare Fig.~\ref{fig:Model_explanation}). This removal rule is based on the assumption that complete capsids are much more stable than partially assembled capsids and that cellular mechanisms exist that are exploited by
the virus to leave the cell. We note that our model does neither incorporate any details of the production mechanism nor any details of the exit mechanism, but is kept as simple as possible in order to investigate the underlying physical principles of the assembly process in such a dynamic setup. For the same reason we focus on the assembly of T1 virus capsids and assume all bonds to be identical. 

For the following it is helpful to introduce the concept of dimensionless bond strength $\ks$, which is defined by the ratio of the microscopic association rate $k_a$ and the microscopic dissociation rate $k_d$:
\begin{align}
    \ks=k_a/k_d\ . \label{eq:bond_strength}
\end{align}
Note that bond strength is similar to, but different from the equilibrium association constant $K_\text{eq}$ for a bimolecular reaction, because it is defined by the ratio of microscopic rates (with the physical dimension $1/s$) rather than by the ratio of a macroscopic association rate constant $k_\text{on}$ (with physical dimension $1/(s M)$) and a macroscopic dissociation rate $k_\text{off}$ (with physical dimension $ 1/s$). 
For the reaction between two clusters (without any closed loops) $K_\text{eq}=V^\star k_s$ is related to $\ks$ by the encounter volume $V^\star$ (with physical dimension $m^3$) which is defined by all two-particle configurations of the two clusters with an overlap of complementary patches \cite{Klein2014}. 
For the case of virus assembly investigated here, $\ks$ is not only a measure for the strength of inter cluster bonds, but also for the stability of closed loops. 

In order to avoid the need to exhaustively scan parameter space and  motivated by previous results quantifying the success of assembly as a function of $k_a$ and $k_d$~\cite{Baschek2012}, we use a linear relation between dissociation and association rate $k_d=m k_a +c$ with $c=\SI{0.0111}{\nano\second} ^{-1}$ and $m=-0.0011$ to explore the parameter space ranging from very strong to very weak bond strength.  When varying  $\ks$ from $10^2$ to $10^5$ we explore association and dissociation rates in the range of $k_a\in [\SI{1}{\nano\second}^{-1},\SI{10}{\nano\second}^{-1}]$ and $k_d \in [\SI{1e-4}{\nano\second}^{-1},\SI{0.01}{\nano\second}^{-1}]$, respectively. 

All simulations have been performed at a time resolution of $\Delta t=\SI{0.01}{\nano\second}$ using periodic boundary conditions. Capsomeres are modeled as hard spheres of radius  $R_\text{steric}=\SI{1}{\nano\meter}$. Each capsomere is equipped with three distinct spherical patches reflecting the geometry of the T1 capsid. Each patch has a radius of $r_\text{patch}=\SI{0.3}{\nano\meter}$ with the center of the patch  being located on the surface of the hard sphere. A bond can only be established between complementary patches according to the local rules. The diffusive properties of all intermediates are represented by their mobility matrices which are evaluated at room temperature $T=\SI{293}{\kelvin}$ using the viscosity of aqueous medium $\eta=\SI{1}{\milli \pascal \second}$. 

Given the typical time and length scales $l_0=\SI{1}{\nano\meter}$ and $t_0=\SI{1}{\nano\second}$ of our simulations, we define the dimensionless length parameter $\lambda=l/l_0$ and the dimensionless time parameter $\tau=t/t_0$ in order to simplify the notation.  Furthermore, we introduce the dimensionless box volume  $\Lambda=V_\text{box}/l_0^3$, the dimensionless particle concentration $\rho=N/\Lambda$, where $N$ is the number of capsomeres in the simulation volume, and the normalized influx rate $\kid=\ki t_0 10^6/\Lambda$. $\kid$ can be understood as the rate of concentration increase due to the influx of capsomeres. Using the normalized influx rate  $\kid$ instead of $\ki$ allows us to compare the assembly process for different sizes of the simulation volume. Our simulation volume has a typical linear extension of 30 to 50 nanometers, leading to reasonable
computing times for complete capsid assembly.

Considering particles which are equipped with only one spherical patch we can relate the values of $\ks$ used here to $K_\text{eq}$ and estimate the speed of the reaction dynamics for this case. In contrast to the capsomeres these particles can only form dimers. The encounter volume for this reaction is $V^\star\approx0.11 {\nano\meter}^3$ (compare reference~\cite{Klein2014} for details on the calculation) and the equilibrium association constants $K_\text{eq}$ range from  $6.6 M^{-1}$ to $6.6\times 10^3 M^{-1}$ for the values of bond strength used here ($10^2 \le k_s \le 10^5$). 
Thus our simulations proceed at a relatively high concentration in the mM-range. 

In order to estimate the influence of the patch size for the speed of the reaction, we use an algorithm developed by Zhou and coworkers~\cite{Zhou1990,Zhou1996} with which the diffusive association rate constant $k_D$ can be estimated based on the survival probability of two clusters starting in an encounter \cite{Klein2014}. For the dimerization we estimate $k_D\approx 6.25 \times 10^8 M^{-1}s^{-1}$.
The diffusive dissociation rate constant then follows as $k_{D,b}=k_D/V^\star$.
Depending on the values of the microscopic reaction rate $k_a$ the macroscopic association rate constant $k_+=k_D k_a/(k_a+k_{D,b})$ is in the range of   $6.0\times10^7 M^{-1} s^{-1}$ to $3.21\times10^8 M^{-1} s^{-1}$. For lower values of $k_+$ the reaction can be considered as reaction-limited ($k_a < k_{D,b}$) while for higher values of $k_+$ the assembly is equally influenced by reaction and diffusion ($k_a\approx k_{D,b}$) \cite{Eigen1974}. Although bimolecular reactions in this range of macroscopic association rate constants $k_+$ are typically considered diffusion-limited \cite{Schreiber2009}, the relatively large patch sizes used here allow for rapid formation of the diffusive encounter so that the reactions in our simulations are at least partly reaction-limited according to the classification scheme by Eigen \cite{Eigen1974}. The macroscopic dissociation rate $k_{-}=k_{D,b}k_d/(k_a+k_{D,b})$ ranges from  $ 4.8 \times{10}^4 s^{-1}$ to $9.0 \times 10^{6} s^{-1}$,  which is approximately the same range as for the microscopic dissociation rate. 

It is instructive to compare the reaction rate constants and equilibrium constants for dimerization used in our simulations to the rates inferred by Xie at al.~\cite{Xie2012} with a non-spatial model from published light-scattering data of the in vitro assembly of hepatitis B virus (HBV) virus \cite{Zlotnick1999}, cowpea chlorotic mottle virus (CCMV) \cite{Zlotnick2000} and human papillomavirus (HPV) \cite{Casini2004}. We first note that
Xie et al. consider closed loops as being infinitely stable which allows for capsid assembly at lower concentrations than in our case.
Nevertheless we find that for HBV and CCMV, our equilibrium association constants are well within the range reported by Xie at al., while our association rate constants are larger by two orders of magnitude. In this context, we note that the use of high concentrations and enhanced assembly dynamics is a common limitation of particle-based simulations of capsid assembly~\cite{Hagan2006,Rapaport2008,Guo2009,Rapaport2010,Baschek2012}.

In our setup, concentrations are not fixed, but arise from the dynamic influx. In practice, 
capsomere and capsid production rates can vary widely for different viruses and different host systems. 
For influenza virus, for example, a production rate of  $10^4$ virions over 10 hours has been reported after start of viral protein translation \cite{Heldt2012}. For HIV, in contrast, only 800 virions are produced over 8 h after start of viral protein translation \cite{Reddy1999}. 
Similar variations also exist for the transcription and translation rates. As we will see below,
our capsid production rate is of the order of $10^8$ virions per hour in a small reaction volume
because we consider a small generic virus with only 60 identical components, relatively high
influx rates and accelerated dynamics.

\section{Results}

\begin{figure}
    \includegraphics{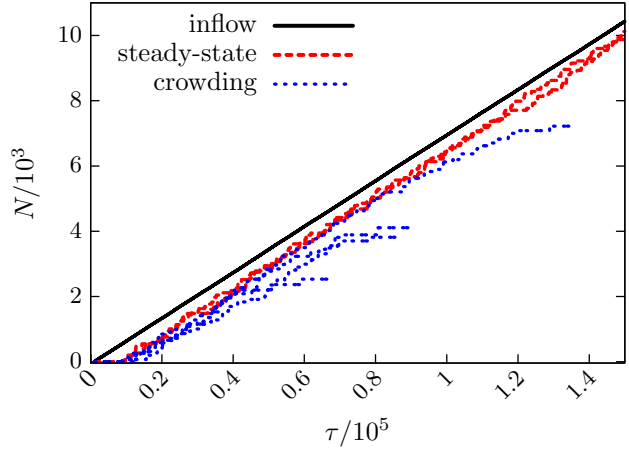}
    \caption{Evolution of the total number of capsomeres placed inside and removed from the simulation volume. The black, solid line depicts the number of capsomeres placed inside the simulation volume with $\kid=2.593$.
    The red, dashed lines show individual trajectories of the number of capsomeres which are removed from the simulation volume due to capsid completion for a bond strength of $\ks=1179$, resulting in a steady state. The blue, dotted lines show individual trajectories of removed capsomeres for a bond strength of $\ks=8483$ at which crowding occurs.}
        \label{fig:inoutflow_vs_time}
\end{figure}

First we qualitatively characterize typical responses of our simulation setup at different bond strengths. For this purpose the time course of the total number of capsomeres placed inside and removed from the simulation volume is shown in Fig.~\ref{fig:inoutflow_vs_time} for two different values of $\ks$ and a normalized influx rate of $\kid=2.593$ (a movie for the inital stages of capsid assembly is
provided as supplementary movie S1). The number of capsomeres placed inside the simulation volume (black, solid line) shows the expected linear increase defined by the influx rate. For the low bond strength of $\ks=1179$ (red, dashed lines) two individual trajectories are shown. For these trajectories we see that after an initial lag phase without capsid completion (the length of the lag phase depends strongly on the rate of capsomere influx and the established concentration in the simulation volume ($0.5-2.0\times \si{10^4}{\nano \second}$)) a steady state with constant capsomere concentration is established in which capsomere influx and capsid removal balance each other  (a movie for the steady state is provided as supplementary movie S2). 

For the higher bond strength of $\ks=8483$ (blue, dotted lines) the trajectories behave very differently. After the initial lag phase the rate of capsomere removal due to capsid completion is almost compensating the capsomere influx for a certain period of time until the rate of capsid completion drastically slows down. The time point at which this slow-down in capsid production is observed strongly varies between different trajectories. Once capsid completion has started to slow down, the concentration in the simulation volume quickly increases due to the influx of further capsomeres and the simulation volume eventually becomes crowded. If the volume fraction of the simulation which is occupied by capsomeres is too large, further assembly is prohibited. For the following we therefore introduce a crowding-threshold. This threshold is reached when 35\% of the simulation volume is occupied by capsomeres and trajectories reaching this threshold are aborted. The chosen threshold is much larger than the highest steady-state concentration established in our simulations and a trajectory reaching this high concentration will certainly lead to a full stop of assembly. 

\begin{figure*}
    \centering
    \includegraphics{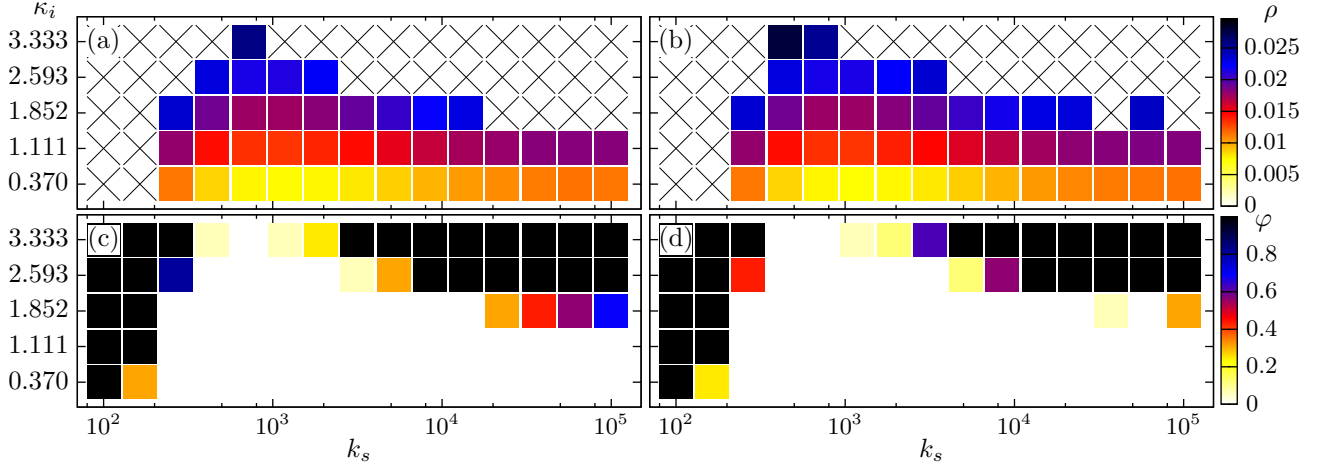}
    \caption{Phase diagrams with the different assembly regimes for different simulation volumes.
    (a) and (b) show the averaged capsomere concentration $\rho$ as a function of influx rate $\kid$ and
    bond strength $\ks$ for a simulation volume of $\Lambda_1=27000$ and $\Lambda_2=42875$, respectively. 
    (c) and (d) show the fraction of trajectories which became crowded during the simulation time as a function of $\kid$ and $\ks$  for a simulation volume of $\Lambda_1$ and $\Lambda_{2}$, respectively.
    Each data point is obtained from 16 independent trajectories using a simulation time of $\tau_\text{sim}=10^6$. 
    }
     \label{fig:combined1}
\end{figure*}
We now systematically investigate the assembly process as a function of bond strength $\ks$ and normalized influx rate $\kid$. In our dynamic setup the rate of capsid production is prescribed by the influx of capsomeres and hence cannot be used as a measure for the quality of the assembly process. Instead we measure the ability of the assembly process to compensate the influx of new capsomeres. As can be seen for the case of a steady state in Fig.~\ref{fig:inoutflow_vs_time} the concentration in the simulation volume linearly increases until a steady concentration is reached at which the assembly of full capsids compensates the influx of new monomers. Thus, for a driven system with a continuous influx of capsomeres the concentration established in the simulation volume for a given $\kid$ is a measure for the efficiency of the assembly process: the lower the established concentration, the higher the efficiency (ability of the assembly process to compensate the influx). 

In Fig.~\ref{fig:combined1} (a) and (b) the average capsomere concentration $\rho$ during our simulations is shown  for two different simulation volumes, respectively. Here only those parameter combinations are shown for which a steady state is established in all trajectories during the simulation time.
If no steady state is established (X), the average concentration does not provide a valid measure for the efficiency of the assembly process as the concentration in trajectories showing crowding rapidly increases until they are finally aborted. Hence for parameter combinations where at least one trajectory shows crowding we instead quantify the ability of the assembly process to compensate the influx of new capsomeres by the fraction of crowded trajectories $\varphi$: the higher the crowding fraction is the smaller the ability of the assembly process to compensate the influx of new capsomeres. This is shown in Fig.~\ref{fig:combined1}(c) and (d) for the two different simulation volumes, respectively.

In general, we see from Fig.~\ref{fig:combined1} (a) and (b) that for every value of $\kid$ used here the lowest concentration is found at intermediate bond strengths of $\ks\approx10^3$. This is similar to assembly under static conditions where intermediate bond strengths show the largest yield of full capsids.
Moreover, a  minimum bond strength of around $2\times 10^2$ is necessary for any assembly to occur. Below this threshold in bond strength the simulation volume becomes crowded almost independent of the normalized influx rate $\kid$. 

For high normalized influx rates  we see that only a narrow range of intermediate bond strengths exists for which a steady state is established in our simulations in all trajectories, and the steady-state concentration even at optimal $\ks$ is very high. Here the simulation volume does not only become crowded for very small $\ks$, but a distinct crowding regime exists at high $\ks$. Thus, in the case of a high forcing of the system due to a rapid influx of capsomeres (corresponding to a fast production of capsids) establishing a steady state requires an optimal bond strength. 

By decreasing the normalized influx rate of capsomeres (lower forcing of the system) the average concentration established in the simulation volume also decreases (Fig.~\ref{fig:combined1} (a) and (b)) as there is more time for the assembly to proceed in between the addition of new capsomeres. Interestingly, the crowding regime at high bond strengths shrinks with decreasing $\kid$ (Fig.~\ref{fig:combined1}(c) and (d)), and for  $\kid=1.111$ a broad range of $\ks$ (also extending to very high $\ks$) exists for which a steady state of capsid production is observed. This is a clear difference between the dynamically driven system analyzed here and static systems with a fixed capsomere concentration. While in the static case kinetic trapping would prevent any capsid assembly at high values of $\ks$, the continuous influx of new capsomeres reduces the requirement for an optimal bond strength and successful assembly is possible even at very high $\ks$. 

As $\kappa_i$ has been normalized by the simulation volume and can be understood as the rate of increase in concentration due to the influx of capsomeres, one expects our findings to be independent of the system size. Comparing the average concentration $\rho$ in the smaller simulation volume (Fig.~\ref{fig:combined1}(a)) and in the larger volume (Fig.~\ref{fig:combined1}(b)) we indeed see that the average concentrations are almost identical in both cases if the rate of capsomere influx is appropriately scaled. This allows for a volume independent comparison of the effect of a dynamic influx of capsomeres on the assembly process. On the contrary, the crowding tendency especially at high bond strengths and high values of $\kid$ is reduced in the larger simulation volume (Fig.~\ref{fig:combined1}d) when compared to the smaller simulation volume (Fig.~\ref{fig:combined1}c). Thus the crowding tendency depends on the system size in a non-trivial manner and reflects the stochastic nature of the crowding process with fluctuations being reduced in larger systems as will be discussed in detail below.

\begin{figure*}
    \includegraphics{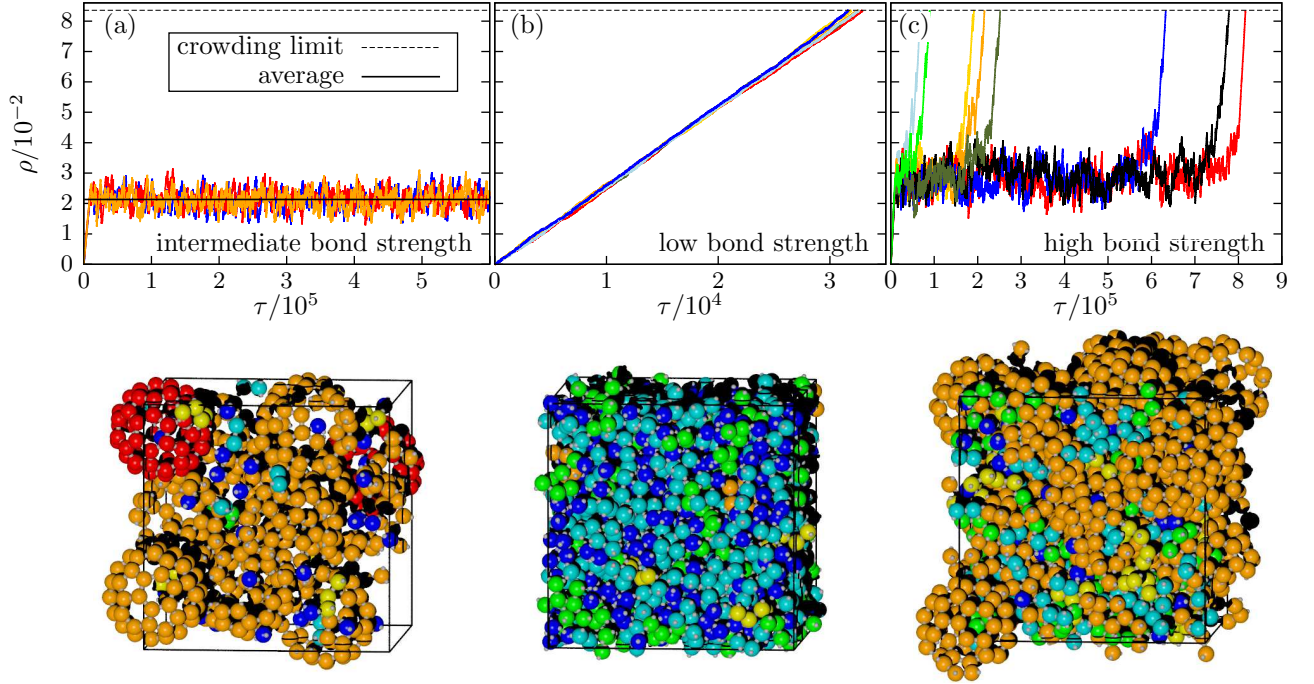}
    \caption{Time evolution of the capsomere concentration for different assembly regimes ($\kid=2.592$, $\Lambda=27000$). For each of the different regimes one representative snapshot is shown in the lower row, with size-dependent color coding. Simulation data for (a) assembly without crowding at intermediate bond strength $\ks=1179$, (b) assembly with crowding at low bond strength $\ks=164$, and (c) assembly with crowding at high bond strength $\ks=8483$. The dashed black line represents the crowding limit at which trajectories are aborted. 
    }
     \label{fig:combined2}
\end{figure*}

In order to further characterize the different regimes, individual trajectories of the time evolution of the capsomere concentration $\rho$ for three different bond strengths with $\kid=2.592$ and $\Lambda=27000$ are shown in Fig.~\ref{fig:combined2} (a)-(c)
(the lower row shows corresponding simulation snapshots).
Fig.~\ref{fig:combined2}(a) shows the evolution of $\rho$ of independent trajectories for intermediate bond strength $\ks=1179$. Here no crowding of the simulation volume occurs and a steady state is established in all simulations. After an initial phase with an increase in $\rho$ a constant concentration is established in all simulations and individual trajectories stochastically fluctuate around this constant concentration. Thus, the average concentration in this case can indeed be used to characterize the system. 

In Fig.~\ref{fig:combined2}(b) the time evolution of $\rho$ for low bond strength $\ks=164$ is shown. In this case the bond strength is below the minimum $\ks$ needed for assembly of larger clusters, and the concentration in all trajectories linearly increases with simulation time leading to an almost deterministic abortion of the simulations due to crowding. 

In Fig.~\ref{fig:combined2}(c) the  evolution of concentration is shown for high bond strength $\ks=8483$. This case corresponds to crowding of the simulation volume at high bond strengths $\ks$ and high normalized influx rate $\kid$. Compared to the almost deterministic crowding process at very low bond strengths (Fig.~\ref{fig:combined2}(b)), we see that the crowding characteristic for this case is fundamentally different as now crowding of the simulation volume occurs stochastically. In all trajectories shown in Fig.~\ref{fig:combined2}(c) a very high, quasi-constant concentration is established for a certain period of time until the system stochastically reaches an unfavorable configuration. Once such a configuration is reached, the formation of further capsids is hindered. This leads to an increase in concentration, which further slows down capsid production presumably due to steric collisions during the reaction process, and the system then quickly becomes crowded with large capsid intermediates due to the addition of new capsomeres. 

A similar crowding characteristic is observed at high $\kid$ for bond strengths ($\ks\approx 268.0$) which are slightly above the threshold in $\ks$ necessary for any capsid assembly.  In this case again capsid completion can balance capsomere influx for a certain period of time until the simulation volume quickly becomes crowded albeit in this case with a high fraction of small clusters. 

In general, stochastic crowding of the simulation volume requires a very high concentration of capsomeres. In this case the realization of an unfavorable configuration hinders capsid assembly for a certain time and the influx of new capsomeres leads to a further increase in concentration which in turn hinders further assembly due to steric collisions. Thus, once a certain concentration is surpassed in our simulations the simulation volume inevitably becomes crowded as the assembly process cannot compensate the influx of new capsomeres. The observation that the crowding process is triggered stochastically by the realization of an unfavorable configuration of the system agrees well with the previous observation that the stochastic crowding tendency decreases with system size (compare Fig.~\ref{fig:combined1} (c) and (d)). For a larger system the relative fluctuations in concentration decrease and the probability that an unfavorable configuration of the whole system is realized is reduced.

\begin{figure}
    \centering
    \includegraphics{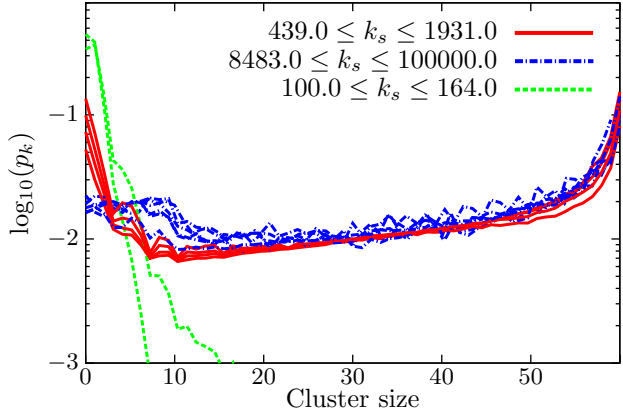}
    \caption{Relative population of different cluster sizes $p_k$ for $\kid=2.593$ and different values of $\ks$ as a function of the cluster size. The red, solid lines correspond to bond strengths without any crowding of the simulation volume ($439\leq \ks \leq 1931$). The blue, dashed and dotted lines and the green, dashed lines correspond to bond strengths where all trajectories showed crowding at high ($\ks\geq8483$) or at low bond strengths ($\ks=100$ and $\ks=164$), respectively.}
        \label{fig:csd_box_crowding_motifs}
\end{figure}
In order to further analyze the mechanisms leading to crowding we compare the relative population of cluster sizes $p_k=f_k \times k/N$ for different cases. Here $f_k$ is the number of clusters of size $k$ and $N$ the total number of capsomeres in our simulation volume. Thus $p_k$ is the probability that an arbitrarily chosen capsomere is part of a k-sized intermediate. In Fig.~\ref{fig:csd_box_crowding_motifs} $p_k$ is shown for different bond strengths and a normalized influx rate of $\kid=2.593$ on a logarithmic scale. Here $p_k$ has been first averaged over the whole simulation time of a trajectory and subsequently over 16 independent trajectories. 

For the intermediate bond strengths for which no crowding is observed (red, solid lines) small and large cluster sizes dominate. This is in agreement with previous observations characterizing successful assembly from a fixed concentration of capsomeres~\cite{Rapaport2004,Zlotnick1994} and indicates that in this regime successful assembly proceeds by the addition of small clusters to only a few larger, stable assembly intermediates. For crowded runs at very low bond strengths (green, dashed lines) only small cluster sizes are populated and no capsid completion is observed. This explains the quasi-deterministic increase in concentration shown in Fig.~\ref{fig:combined2}(b).
For crowded runs at very high bond strengths (blue, dashed and dotted lines) the relative population at large cluster sizes strongly resembles the relative population without crowding. In contrast the population of small cluster sizes is strongly reduced for the crowded runs while the population of intermediate cluster sizes ($k\approx10$) is increased. This shows that in this case  new capsomeres are quickly absorbed by existing clusters and the system becomes eventually crowded with intermediate and large capsid fragments. Interestingly the quick depletion of single capsomeres has previously been identified as characteristic for kinetic trapping during capsid assembly~\cite{Endres2002,Katen2009,Zlotnick2011,Baschek2012}. This shows that kinetic trapping and box crowding at high bond strengths are strongly connected phenomena.

\begin{figure*}
    \centering
    \includegraphics{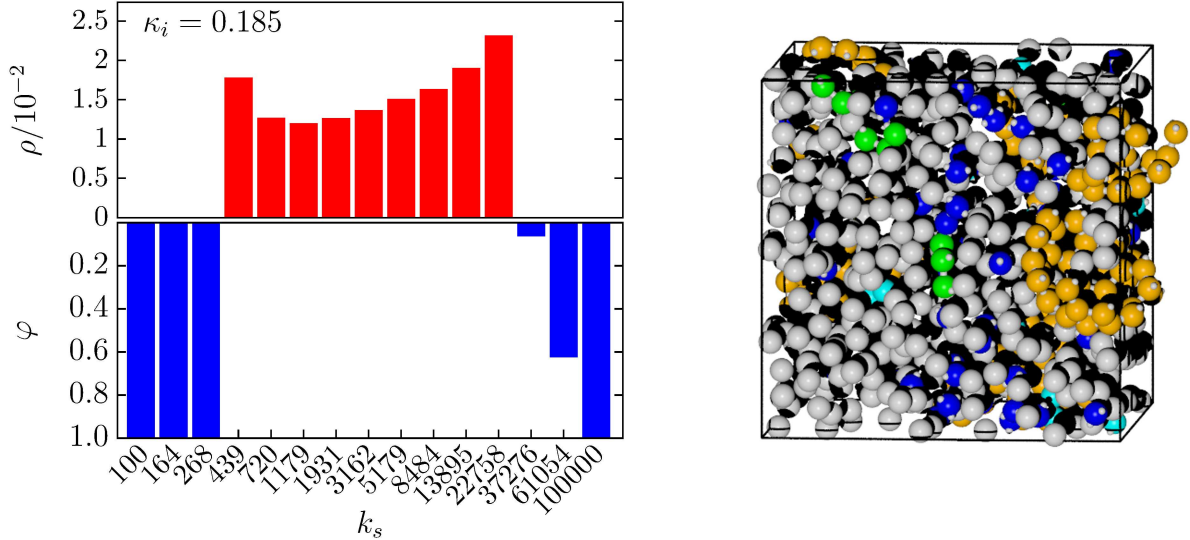}  
    \caption{Averaged capsomere concentration $\rho$ and fraction of crowded trajectories $\varphi$ as a function of bond strength $\ks$ for a normalized capsomere influx rate of $\kappa_i=0.185$ when placing 600 additional, non-specific crowders inside a simulation volume of size $\Lambda_1=27000$.  Each data point is obtained from 16 independent trajectories using a total simulation time of $\tau_\text{sim}=6\times 10^5$. On the right a representative simulation snapshot is shown for $\ks=1179$ and with size-dependent color coding of the capsid fragments. The additional crowders are colored in gray.
    }
     \label{Fig.5_crowders}
\end{figure*}
As we have seen, crowding effects are essential to understand capsid assembly, but until now
we only have considered self-crowding by viral components. In the cell, crowding will be
established also by other crowders and therefore lower (more physiological) concentrations of viral
components are expected to be sufficient to result in similar effects as described here.
In order to test the validity of our findings in the presence of additional, non-specific macromolecular crowders, we have
used our dynamic simulation setup with non-reactive crowders. To this end we place a total of 600 spherical crowders inside a simulations volume of size $\Lambda_1=27000$. These crowders have the same radius as the capsomeres, however, they do not participate in any reactions (no patches). In the simulation volume we now have a total concentration of $\rho_\text{total}=\rho+\rho_\text{crowder}$ with the concentration of crowders $\rho_\text{crowder}$ being kept constant throughout the simulations. This implies that any crowder found inside a complete capsid is placed back into the simulation volume. The upper (red) histogram in Fig.~\ref{Fig.5_crowders} shows the dependence of the average concentration $\rho$ observed during our simulations on the bond strength $\ks$ for a normalized influx rate of $\kappa_i=0.185$. Again $\rho$ is only shown for those values of $\ks$ for which a steady-state was established in all trajectories during the simulation time. The lower (blue) histogram depicts the fraction of crowded runs $\varphi$ for different bond strengths. In addition a representative snapshot of the assembly system including the additional crowding agents is shown in Fig.~\ref{Fig.5_crowders}. Comparing the dependence of $\rho$ on the bond strength $\ks$, we see the same qualitative behavior with additional crowders (Fig.~\ref{Fig.5_crowders}) and without crowders (Fig.~\ref{fig:combined1}). In both cases assembly is most efficient at intermediate bond strengths. Furthermore in Fig.~\ref{Fig.5_crowders} we again observe two regions (at very low and very high bond strength) in which the simulation volume becomes too crowded for a steady state to be established in all trajectories. This is similar to the two self-crowding regions observed in Fig.~\ref{fig:combined1} at high normalized rates of capsomere influx. However, the normalized influx rate of capsomeres used in Fig.~\ref{Fig.5_crowders} is smaller than those used in Fig.~\ref{fig:combined1}. This suggests that, although the qualitative dependence remains the same, it is more difficult to establish a steady-state virus production in already pre-crowded environments and lower rates of influx are required in this case.  

\begin{figure}
   \centering 
   \includegraphics{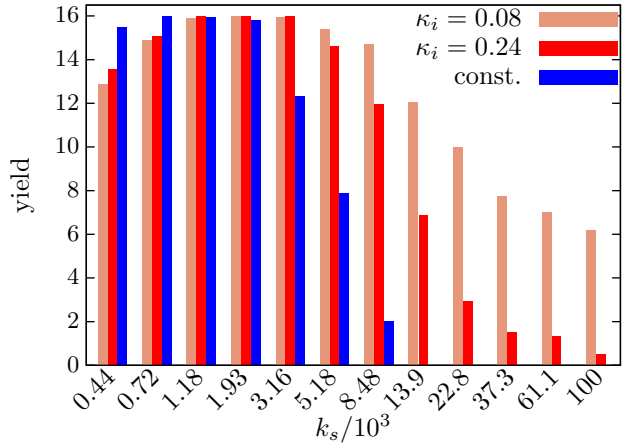}
    \caption{Comparison of capsid yield for a fixed initial capsomere concentration and for a continuous influx of capsomeres. Here the yield of complete capsids in a simulation volume of $\Lambda=125000$ is shown as a function of the bond strength $\ks$ after a simulation time of $\tau_\text{sim}=10^7$. The blue histogram shows the yield when initially placing 1000 capsomeres in the simulation volume. The red histograms on the other hand show the yield of capsids when gradually increasing the number of capsomeres from zero with a rate of $\kid=0.24$ (light red) and $\kid=0.08$ (dark red), respectively, until a total of 1000 capsomeres is reached. The yield has been averaged over 16 different runs.}
        \label{fig:ComparisonInfluxConst}
\end{figure}
After having analyzed  the different mechanisms which prevent a steady state from being established, we now focus on small normalized influx rates. As can be seen in Fig.~\ref{fig:combined1}(a) and (b)
a steady state with continuous capsid production is established even for very high bond strengths if the normalized influx rate is small enough. This suggests that a gradual influx of capsomeres can prevent or at least reduce kinetic trapping. 

In order to verify that dynamic capsomere influx indeed reduces the requirement of an optimal bond strength we compare the yield of capsids for two different setups: a static setup initially starting with $N_0=1000$ capsomeres and a dynamic setup in which a total of $1000$ capsomeres is placed in the simulation volume with a certain rate. In both cases complete capsids are considered to be stable and are taken out of the simulation volume. In Fig.~\ref{fig:ComparisonInfluxConst} the yield of full capsids after a simulation time of $\tau_\text{sim}=10^7$ is shown as a function of the bond strength $\ks$. The blue histogram shows the yield in the static case while the red histograms show the yield when gradually placing new monomers into the simulation volume with a normalized influx rate of $\kid=0.24$ (light red) and $\kid=0.08$ (dark red). 

At low bond strengths ($\ks \leq 720$) the yield in the static case is slightly higher than in the dynamic case. Thus, for small bond strengths it is beneficial if a larger number of capsomeres is available throughout the whole simulation time. In the case of optimal bond strengths ($\ks=1179-1931$) no difference between the two setups is observed as assembly at these bond strengths proceeds quickly, and almost all simulations show the maximum yield of 16 capsids within the simulation time. 

When further increasing the bond strength we see, however, that the yield in the static case quickly drops and no complete capsids are observed above $\ks=8483$. In this case kinetic trapping completely prevents the formation of capsids. For the dynamic case, in contrast, we still observe considerable yield of capsids above $\ks=8483$ with the yield being higher for the lower normalized influx rate used. This indeed shows that a dynamic setup with a gradual supply of capsomeres reduces the selectivity for an optimal bond strength and makes the assembly process more robust and less vulnerable against kinetic trapping. 

\section{Conclusion}

The assembly of the viral protein shell, the capsid,  from elementary assembly units (capsomeres) is a key step during the replication of most viruses. The assembly process needs to be sufficiently robust to guarantee the successful formation of the capsid in the dynamic environment of the host cell. While for test tube experiments on capsid assembly the material available for the assembly process remains constant, in the cellular environment the elementary building blocks for the assembly process are continuously produced by the biomolecular machinery of the cell \cite{Dykeman2014}.

Here we have investigated the role of a dynamic supply of capsomeres and the removal of complete capsids for the assembly of empty T1 virus capsids by using a minimal spatial model based on coarse-grained Brownian dynamics simulations. It has been shown earlier that such a setup can result in a steady state with capsomere influx being balanced by capsid completion \cite{Hagan2011}. Our simulations reveal that for very  high rates of capsomere influx the assembly process is only able to compensate the influx of new capsomeres in a  narrow range of intermediate bond strengths while outside this range the simulation volume becomes crowded. At lower bond strengths the formation of larger clusters is prohibited and the simulation volume becomes crowded almost deterministically with small clusters. At higher bond strengths, in contrast, the simulation volume becomes crowded with large, incompatible assembly intermediates. This crowding process is triggered stochastically by the formation of an unfavorable configuration of the system. While crowding of the simulation volume at very low bond strengths is nearly independent of the influx rate of capsomeres, the crowding regime at high bond strengths vanishes for a slower influx of capsomeres. Thus, for smaller rates of capsomere influx a steady state with continuous capsid production is established even for very high bond strengths.

Recently Smith et al.~\cite{Smith2014} combined a Gillespie type of approach with Green's function reaction dynamics simulations to infer the effect of additional macromolecular crowders on virus capsid assembly. Here we have consider this important aspect in a fully spatial context.
When placing non-reactive macromolecular crowding agents inside our simulation volume, we observe the same trends with regard to bond strength (optimal assembly at intermediate bond strength and crowding of our simulation volume at high and low bond strength).
In the case of extra crowding, our simulations require the use of lower rates of capsomere influx in order to  establish a continuous production of complete capsids, thus bringing our simulations closer to the physiological situation.

Comparing the yield of complete capsids under static conditions with a fixed concentration to the yield of complete capsids when gradually increasing the concentration, we demonstrated that the vulnerability of the assembly process to kinetic trapping can be significantly reduced if the concentration of capsomeres is dynamically increased. This conclusion agrees with the recent finding of a discrete stochastic simulation for genome-stabilized virus assembly that a linear increase in protein concentration dramatically increases the robustness against
kinetic trapping \cite{Dykeman2014}.

To rationalize these results, it is helpful to think of assembly in terms of a free energy landscape (similar to transition networks in protein folding~\cite{Noe2007,Noe2008}). In the picture of the free energy landscape a kinetically trapped assembly process has reached a stable local minimum that prevents the formation of the desired minimum energy configuration. At high bond strengths and high concentrations the free energy landscape of the assembly process is very rough, and trapping in a local minimum is thus very likely. The gradual influx of new assembly material can be understood as a tilting of the free energy landscape. Thus, by continuously providing new capsomeres the assembly process can be guided towards the global minimum corresponding to the formation of full capsids and the dynamic capsomere supply prevents the system from becoming trapped in a local minimum configuration.

As discussed before in our simulations we use enhanced assembly dynamics and relatively large influx rates
to achieve reasonable computing times for our particle-based simulations of empty capsid assembly. 
Thus our simulations include the full effect of diffusional encounters and excluded volume interactions. 
Our simulations suggest that qualitatively similar results are to be expected for lower rates of capsomere influx,
which we use when accounting for the presence of additional macromolecular crowders.
However, further progress in this direction needs algorithmic advances, including the use of GPU-code
and analytical or resampling techniques to speed up simulation times \cite{vanZon2005,vanZon2005_prl,Takahashi2010,Margaret2014}.
Such advances then would allow us to also address more complicated virus architectures with different bond types,
virus misfits, genome-assisted assembly and the interplay between virus assembly and gene expression.

Although our simulations are performed with enhanced assembly dynamics and high rates of capsomeres influx, we believe that our findings have strong implications for the assembly process of virus shells under (dynamic) in vivo conditions. In particular our simulations suggest that kinetic trapping, which is often thought of as a major limitation preventing the successful assembly of complete capsids under (static) in vitro conditions,
might only play a minor role in vivo if the supply of new capsomeres by the host cell is slow enough. Although assembly is still most efficient at intermediate bond strength, a dynamic supply of capsomeres should allow for robust self-assembly in a wide range of bond strengths without the need for additional helper proteins or scaffolds. This finding is compatible with the results by Dykeman et al. \cite{Dykeman2014} and Hagan et al. \cite{Hagan2011} using different setups. Moreover, our dynamic, particle-based simulations show that self-crowding of the simulation volume can prevent steady-steady capsid assembly at high rates of capsomere influx and high bond strengths.

Our simulations suggest that in the presence of other macromolecular crowders the effect of self-crowding due to the continuous production of capsomeres might occur for much lower rates of capsomere influx (and also lower capsomer concentrations). Here further investigations are necessary to clarify the relevance of this regime for in vivo capsid assembly. One way to test our predictions is to use in vitro experiments. The setup studied in our simulations could be experimentally realized using a microfluidic device that allows to control capsomere influx. At the same time, the bond strength might be controlled by changing temperature or ionic conditions. Capsid removal could be implemented simply by sedimentation \cite{Hagan2011}, by filtering or by boundaries that are sticky to completed capsids.  For such a setup,  we expect that the yield of full capsids at high bond strengths (low temperatures) depends on the rate of concentration increase. In particular, decreasing the rate of capsomere influx should result in a higher yield of full capsids. 

As the assembly of a simple icosahedral capsid can be considered as a paradigm for protein assemblies \cite{Johnson1997} or artificial assembly systems in general~\cite{Zhang2003,Olson2007}, our findings also apply for other complex assembly structures. While recent advances in the design of artificial assembly from colloidal particles have aimed at a dynamic control of the inter-particle interactions~\cite{Leunissen2009,Michele2013} to increase the yield of the desired target structure, our simulations strongly suggest that a dynamic control of the material available for the assembly process can further help to increase the yield of the desired structure in artificial self-assembly systems. In these systems the effect of self-crowding discussed in our manuscript might also play an important role, depending on the rate of supply of new assembly material and the accessible assembly volume.

\begin{acknowledgments}
HCRK was supported by a fellowship from
the Cusanuswerk. USS is member of the CellNetworks
cluster of excellence and of the collaborative
research center SFB 1129 (Integrative analysis of
pathogen replication and spread) at Heidelberg.
\end{acknowledgments}

\end{document}